# Long-term analysis of gauge-adjusted radar rainfall accumulations at European scale

Shinju Park*, Marc Berenguer, and Daniel Sempere-Torres

Centre of Applied Research in Hydrometeorology,

Universitat Politècnica de Catalunya-BarcelonaTech, Barcelona, Spain

*Corresponding author address:

Shinju Park, Centre of Applied Research in Hydrometeorology, Universitat Politècnica de Catalunya-BarcelonaTech, Jordi Girona, 1-3, Ed C4-S1, E08034, Barcelona, Spain (shinju.park@crahi.upc.edu).

**Abstract**

Monitoring continental precipitation over Europe with high resolution (2 km, 15 minutes) has been possible since the operational production of the OPERA composites from the European weather radar networks. The OPERA data are the essential input to a hazard assessment tool for identifying localized rainfall-induced flash floods at European scale, and their quality determines the performance of the tool. This paper analyses the OPERA data quality during the warm seasons of 2015-2017 by comparing the estimated rainfall accumulations with the SYNOP rain gauge records over Europe. To compensate the OPERA underestimation, a simple spatially-variable bias adjustment method has been applied. The long-term comparison between the OPERA and gauge point daily rainfall accumulations at the gauge locations shows the benefit of the bias adjustment. Additionally, the daily monitoring shows gradual improvement of the OPERA data year by year. The impact of the quality of the OPERA data for effective flash flood identification is demonstrated for the case of the flash floods that occurred from 29 May to 3 June 2016 in central Europe.

# 1. Introduction

The pan-European radar composites have been operationally produced since early 2012 by the Operational Program for the Exchange of weather RAdar information (OPERA; www.eumetnet.eu/opera) of the Network of European Meteorological Services EUMETNET (Huuskonen et al. 2014). OPERA has been gradually advancing the mosaicking techniques and algorithms for processing heterogeneous radar data from the observations of more than 150 operational radars (Holleman et al. 2006; Scovell et al. 2013; Saltikoff et al. 2018b). The value of the OPERA composites, producing precipitation estimates near ground at continental scale, has increased up to now with applications on the validation and verification of NWP models (Lopez 2014), ground validation of rainfall retrieved from satellite observations (Petković and Kummerow 2014; Beusch et al. 2018), rainfall nowcasting and hydrological forecasting (Berenguer and Sempere-Torres 2013; Sempere-Torres et al. 2016), climate monitoring (Keupp et al. 2017; Saltikoff et al. 2018a) or the study of biological targets (Bauer et al. 2017).

Over the last few years, the OPERA composites have been used particularly as the fundamental input to the European Rainfall-InduCed Hazard Assessment (ERICHA) system developed in the framework of several European projects (HAREN, EDHIT, and ERICHA, see www.ericha.eu). Using the series of the instantaneous OPERA surface rain rate composite, the ERICHA system produces rainfall deterministic nowcasts with lead times up to 6 hours with resolutions of 15 minutes and 2 km (Berenguer and Sempere-Torres 2013) and the flash flood hazard assessment over the 1-km European river (drainage) network (Corral et al. 2019). These products are updated every 15 minutes and have been demonstrated their interest in identifying the most hazardous flash floods in the real-time online ERICHA platform (http://aqua.upc.es/ericha-platform). Since 2017, the ERICHA rainfall nowcasts and flash flood assessment products have been furthermore implemented in the operational European Flood Awareness System (EFAS, http://www.efas.eu; Thielen et al. 2009) to complement the European flash flood forecasts based on a medium-range NWP model with lead times up to 5 days (Raynaud et al. 2014; Smith et al. 2016; Park et al. 2017).

In exploring the use of the OPERA data for the aforementioned applications, some challenges regarding the quality of the OPERA composites (data processed in OPERA 4 from 2013 to 2018) have been reported such as

- Heterogeneity of the radar network, which requires efforts to track timely updates on radar manufacturers, scan strategies, data processing algorithms, and calibration during years among different countries (Peura et al. 2017; Saltikoff et al. 2017).

- Contamination of the OPERA precipitation products due to non-meteorological targets (e.g., ground clutter, presence of birds and insects, and radio interferences; Huuskonen et al. 2014). This has been partially reduced by the OPERA partners (Saltikoff et al. 2018b; e.g., with the implementation of a clutter removal processing that uses a satellite-based cloud mask since late 2015).

- No operational correction applied for neither the vertical profile of reflectivity (VPR) nor path attenuation (Saltikoff et al. 2018b).

- The OPERA rainfall rate is obtained with a single Reflectivity (Z)-Rainfall (R) relationship, i.e., $Z=200R^{1.6}$ assuming the drop size distribution by Marshall and Palmer (1948).

In the near future, further use of the radars with dual-polarization capability should help in improving the quality of the OPERA quantitative precipitation estimates (QPE), similarly as done by Zhang et al. (2016) for the NEXRAD network in the USA.

Rain gauge observations (G) have been widely used as reference to compare and adjust the near-surface radar rainfall estimates (R); e.g., among many authors using single-polarization C-band radar networks in Europe, see Michelson and Koistinen (2000) at multi-national scale; Goudenhoofdt and Delobbe (2009) at national scale; Velasco-Forero et al. (2009) at regional scale. At continental scale, using the OPERA composite, little work has been done on the systematic comparison with gauges (Lopez 2014).

This paper aims at analysing the quality of the OPERA precipitation estimates for the period 2015-2017. The Continental rain gauge records of the WMO surface synoptic observations SYNOP (WMO 2016) have been used as reference. These datasets are described in Section 2. We have also explored the use of a simple real-time gauge adjustment method applied to the OPERA rainfall estimates (Section 3). The results with and without the adjustment between 2015 and 2017 are presented in Section 4. The impact of the gauge-adjusted OPERA data to the flash flood hazard assessment is illustrated through selected events (Section 5). Finally, section 6 summarizes the main conclusions.

## 2. Rainfall accumulations over Europe

ODYSSEY, the OPERA data hub, has been centrally collecting radar volumetric raw data from the EUMETNET member countries and generating three OPERA composites (i.e., instantaneous surface rain rate, instantaneous maximum reflectivity, and one-hour rainfall accumulation; Huuskonen et al., 2014) every 15 minutes. Each composite covers an area of 3,800×4,400 km$^2$ in a Lambert Equal Area projection with 2 km resolution.

In this work, we have used the OPERA instantaneous surface rain-rate composite. These have been processed to produce 15-minute rainfall accumulations, which are the basis of longer-term accumulations. 15-minute accumulations have been computed from 1-minute intermediate frames interpolated accounting for the motion of the precipitation field using the algorithm of Berenguer et al. (2011) and the evolution of precipitation intensities between consecutive composites (see, e.g. Fabry et al 1994).

Figure 1a presents a long-term accumulation map generated with the OPERA composites for the period from 26 January to 24 July 2017 (180 days). This accumulation shows the spatial variability of the precipitation accumulations at Continental scale. However, it also reveals the presence of some of the radar artefacts affecting the radar rainfall estimates; e.g., i) terrain interception (Romanian mountain ranges between 20° E and 30° E longitude, around 47° N latitude), ii) sea surface reflection (in some locations near the Mediterranean coast of France), iii) orographic beam blockages (Northern coast of the Majorca island between 0° E and 5° E longitude, around 40° N latitude, in the interior of Norway, in the Alps), iv) severe radio interferences by wireless networks or radio links pointing toward the radar locations (Southeast of the Iberian Peninsula), or iv) unknown sources (ring patterns in the centre of the domain).

The accumulation for the same period estimated from the available SYNOP precipitation records is shown in Fig. 1b. Although this gauge network is sparse in some countries (especially in comparison with some of the existing national or regional networks), its coverage extends over Europe and its records are freely available from the OGIMET website (www.ogimet.com). Because the precipitation records are obtained with different accumulation windows (e.g.,1, 3, 6, 12 or 24 hours depending on the daily conditions of each ground station), we have homogenized and accumulated them in a common 24-h window (from 0000 to 2400 UTC). Note that the gauge data can be also erroneous in the presence of strong winds, snow, or damages on the devices (Kidd et al., 2017). To avoid the most obvious errors, we have systematically filtered the gauge data out by imposing some coherence with the collocated OPERA estimates. The filtering is done using three criteria to define valid radar- rain gauge (R-G) pairs: (i) areas not affected by ground clutter (a static mask, grey areas in Fig. 1a, has been retrieved to label the areas where both radar echoes are unrealistically too frequent and estimated accumulations are systematically high), (ii) the minimum and maximum values of 24-h rainfall accumulation (a threshold of 1 mm is used to define rainy days, and accumulations over 300 mm are considered not realistic), (iii) the maximum allowed radar-rain gauge difference (if the difference exceeds 90 mm, the radar-rain gauge pair is considered erroneous). These criteria are arbitrary, but they have been useful to guarantee a minimum quality of the rain gauge records used in this study.

Although the spatial distribution of the rainfall accumulations of Figs. 1a and 1b shows clear correspondence, the OPERA estimates over most areas are clearly lower than the interpolated gauge estimates. These significant biases suggest the need for adjusting the OPERA estimates with the available rain gauge observations.

If the gauge data were available in real-time with short delay and at high temporal resolution (e.g. hourly), this adjustment could be done using radar-raingauge blending techniques (e.g., Velasco-Forero et al. 2009; Zhang et al. 2016). However, given that the European-wide SYNOP data are available only at the end of the day, in this work we use a bias-adjustment method based on long-term comparison between radar and rain gauge accumulations to monitor and quantify the systematic biases throughout the OPERA coverage.

## 3. A real-time bias adjustment method

The proposed method to adjust the OPERA precipitation estimates is based on finding a systematic bias throughout Europe so that the adjusted precipitation can be closer to the gauge amounts. The method is similar to Brandes (1975), and it is adapted to the following constraints of data availability and requirements: i) coarse temporal resolution of rain gauge observations (24 hours), ii) availability of the rain gauge data at the end of the day that leads to the resulting bias to be updated once a day, iii) adaptation to real-time performance, and iv) simplicity.

First, daily rainfall accumulations at the location of the $i$-th gauge are estimated from the OPERA composites, $R_i$, and from the SYNOP gauge records, $G_i$. At a given rain gauge location, $(x_i, y_i)$, a multiplicative bias is estimated as the ratio between the radar and rain gauge amounts:

$$B(x_i, y_i) = \frac{<R_i>_t}{<G_i>_t} \qquad [1]$$

The bracket $<\ >$ denotes the total precipitation accumulated in the last $t$ valid rainy days (which in this study a day is classified as rainy when both $R_i$ and $G_i$ exceed 1 mm, as mentioned in Section 2). To find the most recent $t$ rainy days (not necessarily to be consecutive), the method analyses daily rainfall starting from the previous day up to 180 days backward and uses the rain gauge location where the collocated time series of 24-h accumulation show a Pearson correlation larger than 0.60.

To compensate the biases, at the valid rain gauge location, radar rainfall estimates can be adjusted with a multiplicative factor,

$$F(x_i, y_i) = \frac{1}{B(x_i, y_i)} = \frac{<G_i>_t}{<R_i>_t} \qquad [2]$$

A map of the adjustment factor in logarithmic units [dB(R)] over the OPERA grid, $F(x, y)$, is then computed by the inverse distance weighted interpolation of the values at rain gauge locations with Eq. [3].

$$10 \log [F(x,y)] = \frac{1}{\sum_{i=1}^{n} w_i(x,y)} \sum_{i=1}^{n} w_i(x,y) 10 \log \left( \frac{<G_i>_t}{<R_i>_t} \right) \quad [3]$$

Where *n* is the number of rain gauges, and the weights are inversely proportional to the distance, $d_i$; i.e. $w_i(x,y) = \frac{1}{d_i(x,y)}$. All the data points available within a distance of 160 km have been used in the interpolation, ensuring a minimum number of 6 points in the areas with low rain gauge density.

Note that the value of *t* in Eq. [3] is set to seven days considering daily rain gauge data availability (see Fig. 2a indicated by the mean number of days), mean number of rainy days (Fig. 2b), and the presence of long-lasting dry periods (Fig. 2c indicated by the maximum consecutive dry days) from May to October during 2015-2017. For instance, in the Iberian Peninsula, the mean number of days with accumulations exceeding 1 mm is as small as seven days (Fig. 2b), and this area experienced no precipitation for more than three months consecutively (Fig. 2c). Hence, setting such a threshold can gurantee both sufficient rain and representativeness in space with enough gauges reporting rain (larger values of $t$ would discard many rain gauges in the Southern part of the Iberian Peninsula, where the number of rainy days is relatively small).

To obtain the gauge-adjusted values $R_a(x,y)$, the interpolated adjustment factor $F(x,y)$ is applied to the OPERA instantaneous rainfall intensities, $R$, as

$$R_a(x,y) = F(x,y)R(x,y) \quad [4]$$

Figure 3 shows an example corresponding to the case of 25 July 2017 when heavy rainfall occurred over Central Europe (northwestern Poland, Northeastern and Central Germany) and triggered floods in some of these areas the following day. The OPERA 24-h accumulations over these areas (Fig. 3a) are clearly underestimated compared to the interpolated estimates from the SYNOP gauges (Fig. 3b). In Fig. 3c, the map of adjustment factor, $F(x,y)$ computed with Eq. [3], varies smoothly in space. Most land areas show factors between 1.2 and 2.6. High values appear in the areas mostly affected by some poor radar coverages (e.g., affected by mountain blockages or at the edge of radar coverage). In these areas, the maximum value of the adjustment factor is set to 5. After applying the adjustment factor to the original OPERA rainfall intensity maps according to Eq. [4] in real time (every 15 minutes on 25 July 2017), the resulting daily accumulation map (Fig. 3d) becomes more similar to the gauge-based estimates, while providing much more spatial detail. This can also be seen in Fig. 4 with scatter plots between the valid daily gauge records (Fig. 3b) and the collocated OPERA estimates before and after the use of the adjustment. After the gauge adjustment (Fig. 4b), the R/G ratio becomes closer to 1 indicating the adjusted radar amounts agree more to the gauge amounts. Other skill scores (e.g., higher Pearson correlation coefficient, lower mean absolute error, lower root mean square error) also indicate better performance after the adjustment.

## 4. Long-term analyses of gauge-adjusted radar accumulation

To validate the robustness of the method, we have extended the daily accumulation point comparisons among OPERA, gauge-adjusted OPERA, and the SYNOP gauges to a longer period. Warm months (01 May to 31 October) from 2015 to 2017 are chosen to focus on the OPERA rainfall estimates and avoid snow cases as much as possible for which the use of Z-R relationship introduces large errors (e.g., Lopez 2014). First, the time series of the ratio R/G before and after the gauge adjustment are presented in Fig. 5 (blueish and reddish lines, respectively). We can clearly see that a serious systematic underestimation of the OPERA daily accumulations improves in all years after the adjustment; i.e., the values of daily R/G ratio below 0.7 (see Table 1, median 0.48, 0.5, and 0.61 for the periods in 2015, 2016, and 2017, respectively) become between 0.7 and 1.3 (median 1.04, 1.05 and 1.06). The results are not affected by the different sampling number of rainy gauges (shown with different symbols). Note that, the R/G values before the adjustment in 2017 (Fig. 5c) are still below 1 but slightly closer to 1 than the previous years (Fig. 5a and 5b), which is possibly related to some improvements or changes in the OPERA data (e.g., better hardware monitoring from the upgrade of national radars, data sent to ODYSSEY, and the OPERA quality processing) throughout 2017 (Saltikoff et al. 2018b).

Similarly, the correlation coefficient values (CORR, Fig. 6) between the gauge-adjusted OPERA and gauge daily accumulations are higher than those without the adjustment for most days, indicating the stable performance of the adjustment factor. The sudden drop of correlation coefficient values in a limited number of days is possibly caused by the presence of a small number of valid gauge accumulations (e.g., less than 100 gauges recorded little amount of rain over the Continent on 01 July 2015, 26 September 2015, 13 August 2016, and 7 October 2016), or some missing radar observations in the OPERA composites (01 May 2016).

Note that the time series of daily R-G correlation coefficient is less fluctuating for the OPERA data without the gauge adjustment during 2017, compared to the previous years (see Table 1, the decreasing normalized root mean square error year after year at 15, 50, 85 percentiles). It is possibly due to the effect of gradual changes in the OPERA data aforementioned. Similar results are also supported by the time series of the mean absolute error (MAE, Fig. 7 and Table 1). The MAE values are reduced after applying the gauge adjustments and better in 2017 than the previous years except for a few days, where the dominant precipitations occur over the Alps or Pyrenees (e.g., 22 May or 14 June 2017), beam blocked areas, or the edge of the OPERA coverage (e.g., 26 May 2017).

## 5. Impact on the flash flood identification

Would this simple real-time gauge adjustment of OPERA QPE lead to better results in the flash flood hazard assessment? To answer this, we have selected a case to illustrate the use of gauge-adjusted OPERA composite for the pan-European flash flood hazard assessment in basins between 10-5000 km² with the ERICHA flash flood approach (ERICHA-FF; Park et al. 2017; Corral et al. 2019). This approach is built using the 15-minutes OPERA rainfall accumulations as inputs. The radar rainfall aggregated over the upstream basin for an accumulation window corresponding to the concentration time (i.e., called basin-aggregated rainfall) is computed as a proxy to characterize the potential flash flood hazard (Corral et al. 2009). Every 15 minutes, the hazard level (low-yellow, medium-orange, and high-red) is estimated at each point of the drainage network at European scale (with a resolution of 1 km) according to a set of rainfall Intensity-Duration thresholds for the basin-aggregated rainfall. This is adapted from the thresholds used by METEOALARM (www.meteoalarm.eu) developed through the EUMETNET-EMMA project. The ERICHA-FF hazard assessment is solely based on rainfall inputs assuming that the basin response will be strongly dominated by the rainfall. This is particularly valid for high-return period events, and it simplifies the complexity of hydrologic and hydraulic processes (e.g., the initial moisture state, ground water deficit, snow cover and melt, land surface temperature, vegetation, and routing) and can be useful in ungauged basins (Alfieri et al. 2015). These are all relevant advantages for an operational implementation over large domains (in this case, Europe), where the aim of the system is to detect flash flood events in small and medium basins that are not often gauged.

Clearly, this simple identification of flash floods depends critically on the quality of the rainfall inputs and particularly on the estimated amounts. Figure 8 shows an example of the ERICHA-FF hazard level identified during the heavy rain events that resulted in flash floods several parts in central Europe from 29 May to 3 June 2016 (Fig 8a). Based on the real-time OPERA precipitation composites, the total rainfall accumulation during the period is presented with and without the gauge adjustment in Figs. 8b. and 8c, respectively. This helps to quickly identify the areas affected by large rain accumulations (e.g., over 50 mm), which resulted from i) fast-moving (short-lived) individual convective storms and ii) several mesoscale convective systems moving from east to west (not shown here but visible in the animation of 15-min OPERA rainfall accumulations). In Fig. 8, the cyan ellipses indicate the areas where some floods were reported in the news day after day: i) Braunsbach and Schwäbisch Gmünd[1], Germany, on 29 May (Bronstert et al. 2018), ii) Antwerp[2], Belgium, on 30 May, and iii) Simbach am Inn[3], Germany, on 01-02 June (Piper et al. 2016). Figures 8b and 8c show that the 6-days rainfall accumulation increases significantly after the gauge adjustment over the areas mentioned above; respectively i) from 30-60 mm to 60-110 mm, ii) from 30 mm to 50 mm, and iii) from 50-70 mm to 120-160 mm. Consequently, the performance of the ERICHA flash flood hazard level (updated every 15 minutes), summarized in terms of the maximum hazard level estimated for the six days (Figs. 8d and 8e), is also affected; i.e., the input of the real-time gauge-adjusted OPERA rainfall has resulted in many more points of the drainage network with significant hazard levels. The hazard levels also show a wider range, in general becoming higher. The medium and high hazard levels correspond better to the damage points reported in the European Severe Weather Database (ESWD; www.eswd.eu, Dotzek et al. 2009a) for the same days (overlain in Fig. 8f), and to the discharge simulations of Bronstert et al. (2018), which show high return periods for the case of Braunsbach and Schwäbisch Gmünd.

---

[1] http://floodlist.com/europe/germany-floods-baden-wurttemberg-may-2016

[2] https://www.vrt.be/vrtnws/en/2016/05/30/heavy_rain_causingproblemsacrossflanders-1-2669674/

[3] http://floodlist.com/europe/germany-deadly-floods-hit-bavaria-june-2016

Therefore, the gauge adjustment to the OPERA composite (at least collected during OPERA 4 from 2013 to 2018) was crucial to the application of the flash flood hazard assessment.

## 6. Conclusion

The Continental precipitation estimates from the European radar networks (i.e., the EUMETNET-OPERA composites) are the main input for several applications, and particularly for the flash flood hazard assessment and forecasting. These applications depend on the input quality that is fundamental to be monitored and guaranteed. The significant biases found in the OPERA estimates suggested to apply a simple adjustment factor to compensate the rainfall underestimation using European-wide rain gauge observations. Although the applied method is very simple and is not new, it is adapted to the real-time conditions given the rain gauge observations available throughout Europe. Additionally, the daily estimated bias map serves as an indicator of the quality of the OPERA data and has been used to provide feedback to the OPERA community.

The results presented here analyse for the first time the evolution of the quality of the OPERA rainfall estimates with the SYNOP gauge adjustment over the warm seasons of 2015-2017. The systematic differences between OPERA and rain gauge daily accumulations could be reduced significantly after applying the adjustment factor. This resulted in R/G values and correlation coefficients closer to 1, and smaller errors compared with those without applying the adjustment. The daily monitoring of those skill scores also suggests that the OPERA data itself has been gradually improved. This can be seen in Fig. 9 and Table 1 with more recent results obtained from 01 May to 30 Sep 2018 (e.g., the values of median R/G closer to 1: 0.48, 0.5, 0.61, and 0.89 for the periods in 2015, 2016, 2017, and 2018 respectively). The reason for such clear improvements in the R/G and MAE values of the non-adjusted OPERA products is not known (beyond the scope of this work) but can be the result of some changes in the OPERA processing (e.g., the collection of the best possible data implemented by 11 members by May 2018, Saltikoff et al. 2018b). On the other hand, the change of the input OPERA data might result in a slight overestimation after the gauge-adjustment (e.g., from May to July 2018 in Fig. 9a and slightly increased quantile values of R/G, MAE, CORR, and RMSE). Because the adjustment method assumes the biases to remain reasonably constant over a certain time (the seven most recent rainy days over each rain gauge), it will require some time to adjust to the change. Nevertheless, the quantile values show that the overall performance of the gauge adjustment is stable compared to the previous years and similar to those obtained for OPERA (without the gauge adjustment) during the warm season of 2018.

The use of the gauge-adjusted OPERA rainfall composites has clearly shown a positive impact on the performance of a flash flood hazard assessment (ERICHA-FF), producing more realistic flash flood hazard levels over the areas affected by heavy rainfalls in Central Europe from 29 May to 3 June 2016. More importantly, the result shows the usefulness of the OPERA data in the continental assessment of flash floods triggered by locally heavy rainfall, which can support existing hazard forecasting systems (for instance at national scale or relying on NWP forecasts; see, Alfieri et al., 2012; Emerton et al. 2016; Corral et al., 2019). However, there are still data quality issues addressed in the introduction that affect the performance of the ERICHA flash flood indicator. Ultimately, constant improvement foreseen in the processing of the OPERA products from the volume scans of individual radars (including the removal of artefacts -caused by ground and sea clutter, beam blockage or wireless interferences-, attenuation and VPR correction, and the use of dual-polarization) will have a direct impact on the quantitative applications of the continental radar composites.

**Acknowledgements**

The authors are very grateful to Dr. Elena Saltikoff (FMI, the OPERA project manager during 2013-2019) for her encouragement during the work. This work has been carried out within the European Union DG ECHO Prevention Projects ERICHA (ECHO-SUB-2015-718684) and SMUFF (UCPM-2017-PP-PREV-AG-783237), and H2020 Project ANYWHERE (H2020-DRS-1-2015-700099).

Table 1. Summary of the evaluation of OPERA daily rainfall accumulations without and with the gauge adjustment. Three quantiles (15, 50, and 85 percentiles) during warm seasons from 2015 to 2018 are computed for five skill scores (some presented in Figs. 5, 6, and 7): i.e., R/G ratio (1 indicates there is no bias between radar and gauge estimates), mean absolute error (MAE, the smaller the better), and correlation coefficient (CORR). Quantiles of both root mean square error (RMSE) and normalized root mean square error are added though the daily time series are not shown here. The normalized RMSE is defined as the ratio between the daily RMSE and the mean of daily gauge accumulation at gauge locations (smaller values indicating less spatial variability of the accumulated field). The scores are calculated using the valid daily R-G pairs.

| Skill score quantile [%] | | | 2015 May-Oct | 2016 May-Oct | 2017 May-Oct | 2018 May-Sep |
|---|---|---|---|---|---|---|
| R/G | Non-adjusted | 15 | 0.38 | 0.39 | 0.51 | 0.68 |
| | | **50** | **0.48** | **0.50** | **0.61** | **0.89** |
| | | 85 | 0.60 | 0.58 | 0.75 | 1.06 |
| | Gauge-adjusted | 15 | 0.89 | 0.91 | 0.89 | 0.93 |
| | | **50** | **1.04** | **1.05** | **1.06** | **1.11** |
| | | 85 | 1.27 | 1.23 | 1.26 | 1.31 |
| MAE | Non-adjusted | 15 | 3.59 | 3.56 | 3.27 | 3.41 |
| | | **50** | **5.00** | **5.16** | **4.26** | **4.05** |
| | | 85 | 7.18 | 7.04 | 5.61 | 5.03 |
| | Gauge-adjusted | 15 | 3.25 | 3.50 | 3.28 | 3.42 |
| | | **50** | **4.31** | **4.36** | **3.88** | **4.31** |
| | | 85 | 5.50 | 5.61 | 4.95 | 5.04 |
| CORR | Non-adjusted | 15 | 0.40 | 0.40 | 0.49 | 0.53 |
| | | **50** | **0.53** | **0.56** | **0.64** | **0.67** |
| | | 85 | 0.65 | 0.68 | 0.73 | 0.76 |
| | Gauge-adjusted | 15 | 0.49 | 0.52 | 0.58 | 0.58 |
| | | **50** | **0.60** | **0.65** | **0.69** | **0.69** |
| | | 85 | 0.72 | 0.74 | 0.77 | 0.78 |
| RMSE | Non-adjusted | 15 | 5.49 | 5.62 | 5.30 | 5.32 |
| | | **50** | **7.76** | **7.94** | **6.93** | **6.44** |
| | | 85 | 11.71 | 11.28 | 8.97 | 8.02 |
| | Gauge-adjusted | 15 | 4.97 | 4.96 | 4.95 | 5.31 |
| | | **50** | **6.60** | **6.71** | **6.12** | **6.53** |
| | | 85 | 9.13 | 8.67 | 8.04 | 7.72 |
| Normalized RMSE | Non-adjusted | 15 | 0.84 | 0.83 | 0.75 | 0.71 |
| | | **50** | **0.95** | **0.91** | **0.85** | **0.82** |
| | | 85 | 1.09 | 1.06 | 1.00 | 0.96 |
| | Gauge-adjusted | 15 | 0.68 | 0.64 | 0.62 | 0.68 |
| | | **50** | **0.80** | **0.76** | **0.77** | **0.80** |
| | | 85 | 0.94 | 0.92 | 0.98 | 0.98 |

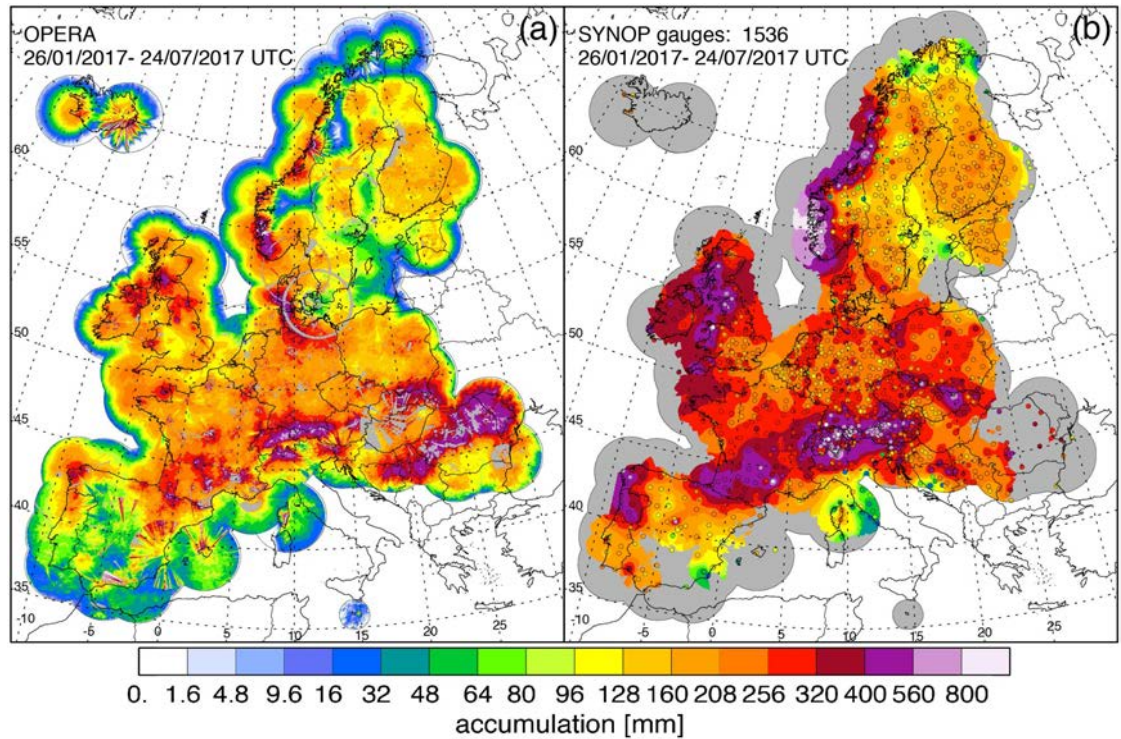

Fig. 1: 180-days precipitation accumulation (from 26 January to 24 July 2017) obtained from (a) the 15-min accumulations estimated from OPERA composites with the spatial resolution of 2 km (the grey areas indicate the areas of radar clutter identified during dry periods in 2015), and (b) the SYNOP-gauges records: The values obtained from the valid rain gauges 1536 (circles) out of the 2629 within the OPERA coverage (not shown) are interpolated to the radar grid by the inverse distance weighted interpolation considering at least six rain gauges and as many gauges available within a maximum distance of 160 km (the areas where not enough rain gauges were available are shown in grey).

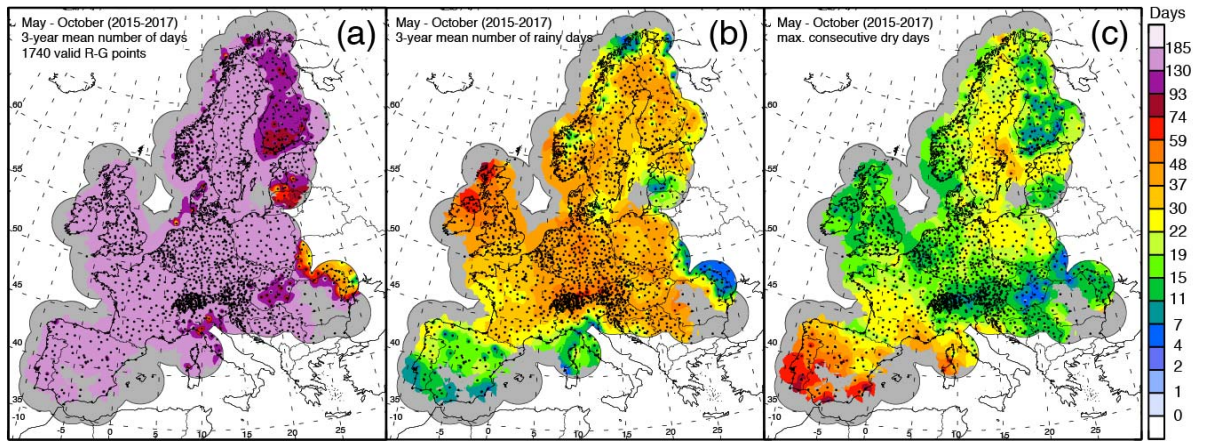

Fig. 2: (a) Mean number of days with valid G records (including both dry and rainy days), (b) mean number of rainy days, and (c) maximum consecutive dry days from May to October during 2015-2017. The black dots indicate the location of the gauges that recorded rain at least one day during the analysed period (a total number of 1740), and the interpolation of resulting days at each gauge done similarly as in Fig.1.

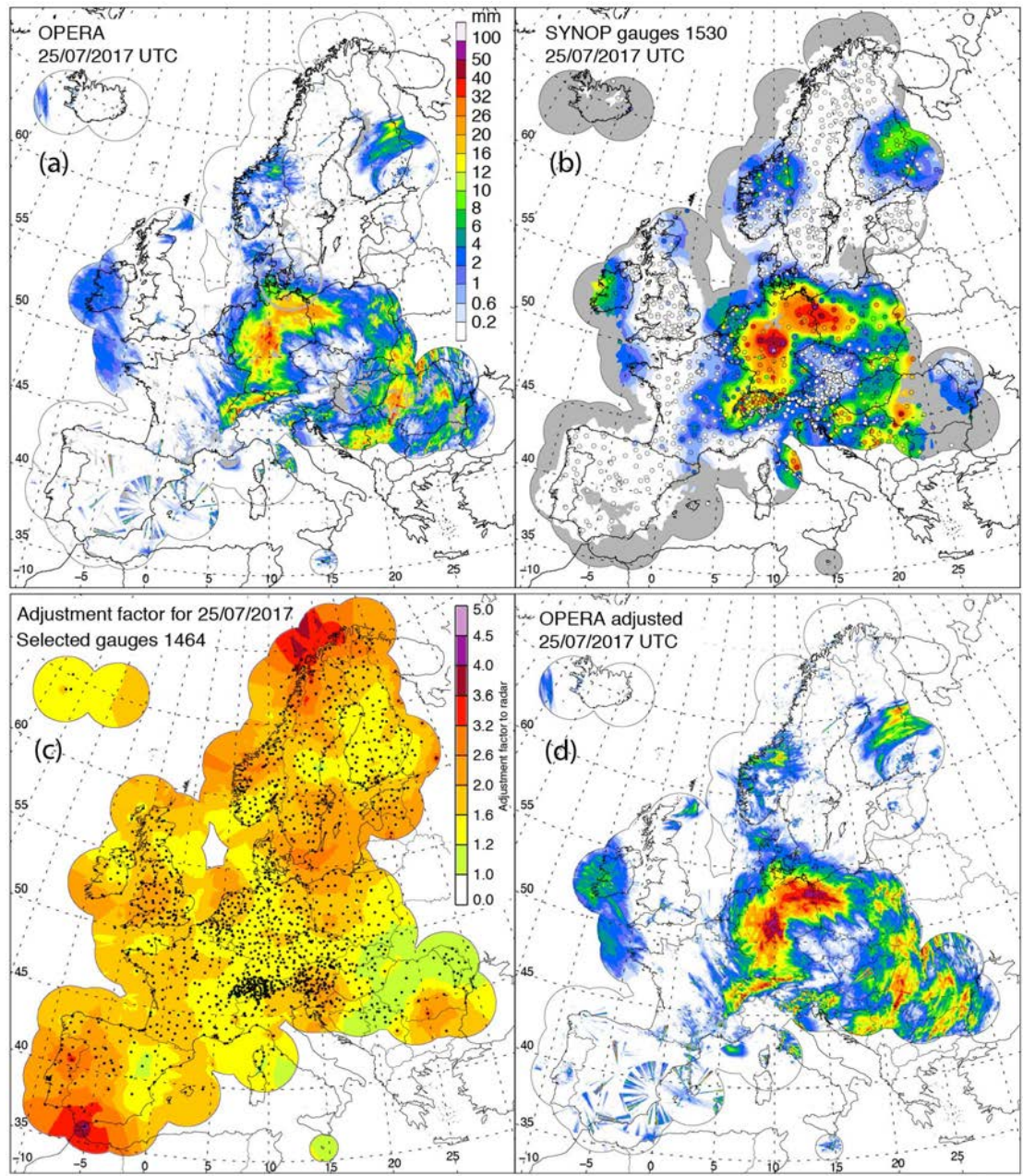

Fig. 3: An example of the gauge adjustment factor applied to the real-time OPERA composites on 25 July 2017: (a) OPERA 24-h rain accumulation before the gauge adjustment. (b) 24-h accumulation estimated by the inverse distance weighted interpolation of SYNOP gauges; 1530 rain gauges were used, out of which 511 recorded accumulations over 1mm (the latter have been used for validation; see Fig. 4); grey areas here indicate no values are recorded within the OPERA coverage. (c) the map of gauge adjustment factor interpolated with the point differences between OPERA and SYNOP gauge accumulations estimated at the valid gauge locations (a total number of 1464, black dots, constrained by the most recent 7 rainy days) during the period from 26 January to 24 July 2017 (see Eq. [3]). (d) the gauge-adjusted OPERA 24-h rain accumulation.

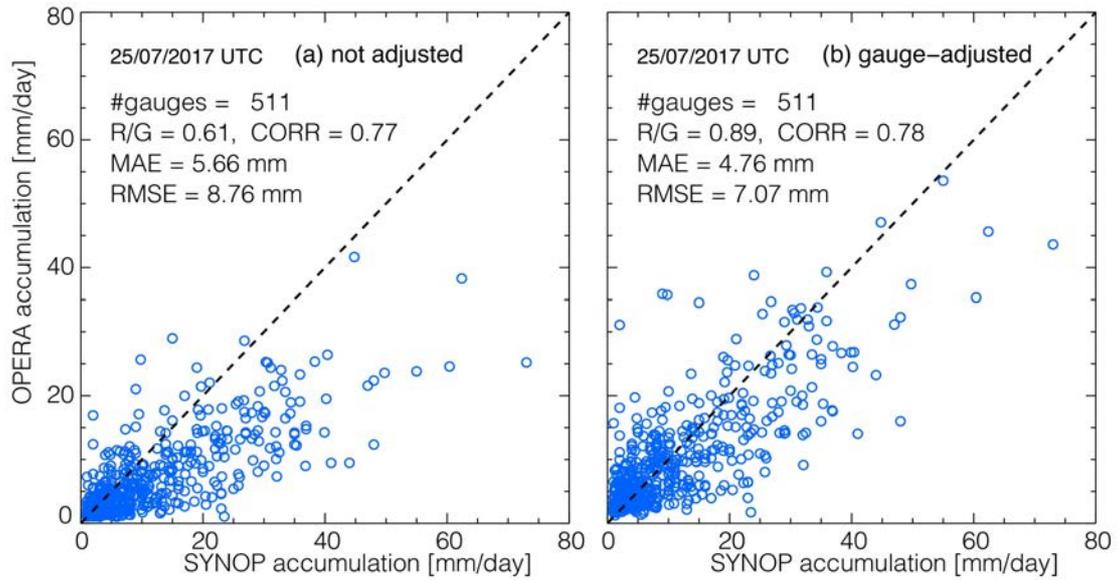

Fig. 4: Scatter plots between OPERA radar and rain gauge daily accumulation obtained at the valid gauge locations (total number of 511) seen in Fig. 3b (25 July 2017). The ratio (R: Radar, G: gauge), mean absolute error (MAE), root mean square error (RMSE), correlation (CORR) are computed (a) before and (b) after the gauge adjustment.

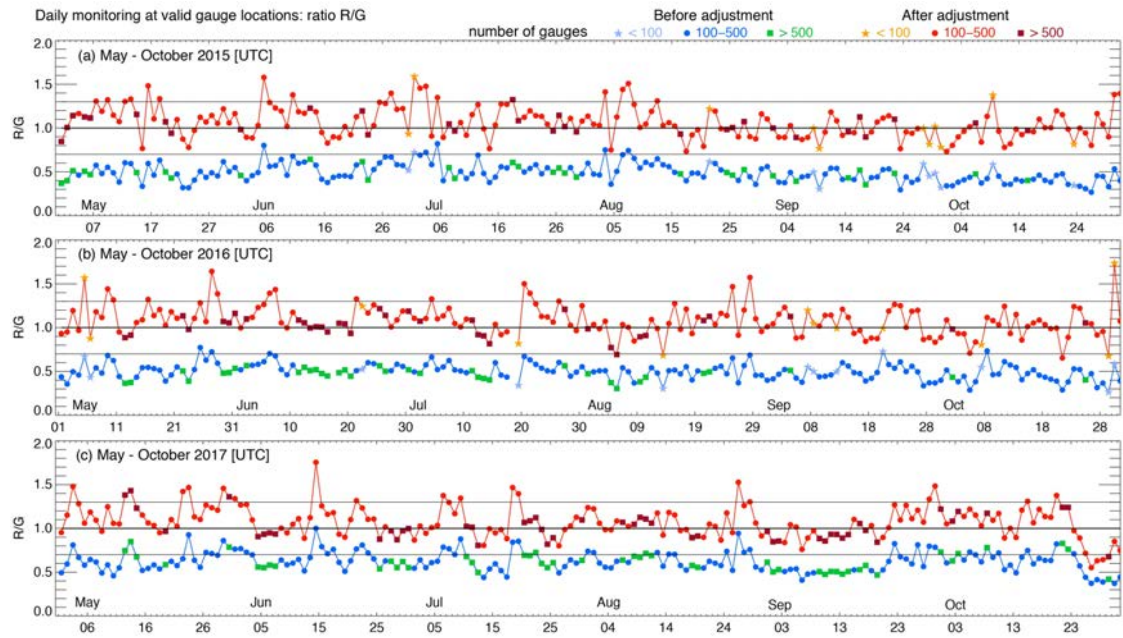

Fig. 5: Daily monitoring of the ratio (R: radar-OPERA, G: gauge) before (blueish colours) and after (reddish colours) the gauge adjustment from May to October in (a) 2015, (b) 2016, and (c) 2017. The number of valid gauges for the comparison varies and is displayed with slightly different colours and symbols.

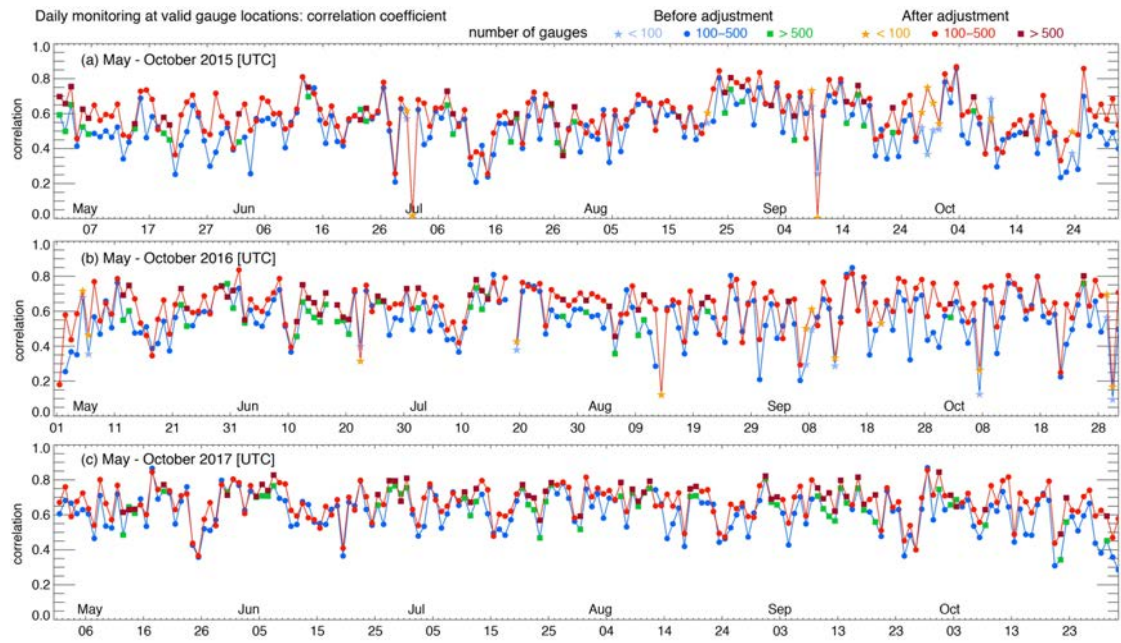

Fig. 6: Similar to Fig. 5 but the daily monitoring of correlation coefficient.

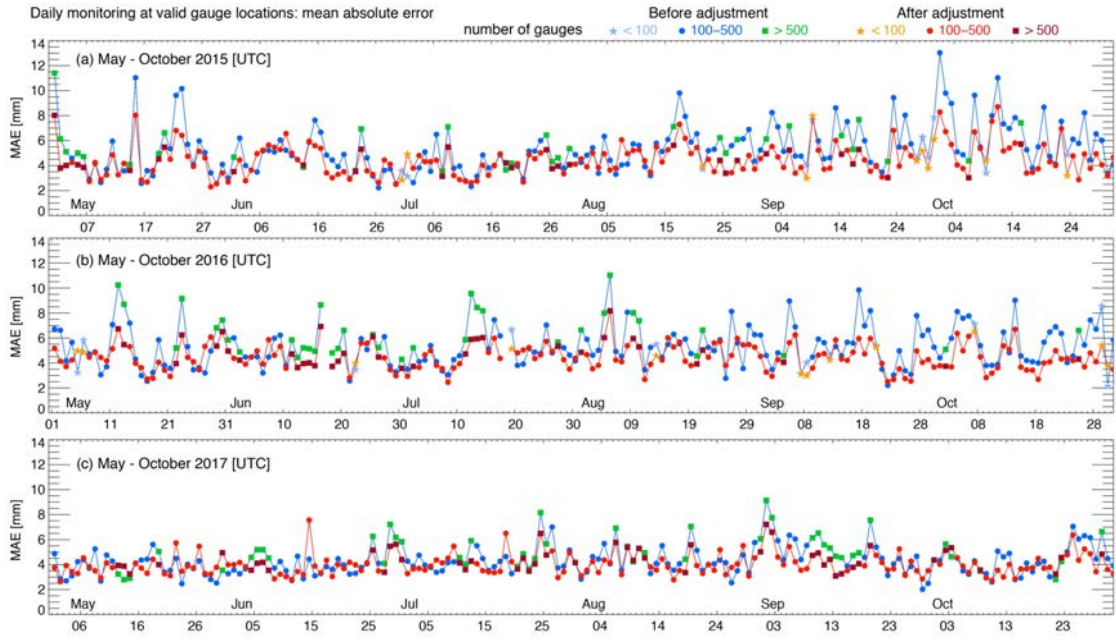

Fig. 7: Similar to Fig. 5 but the daily monitoring of mean absolute error.

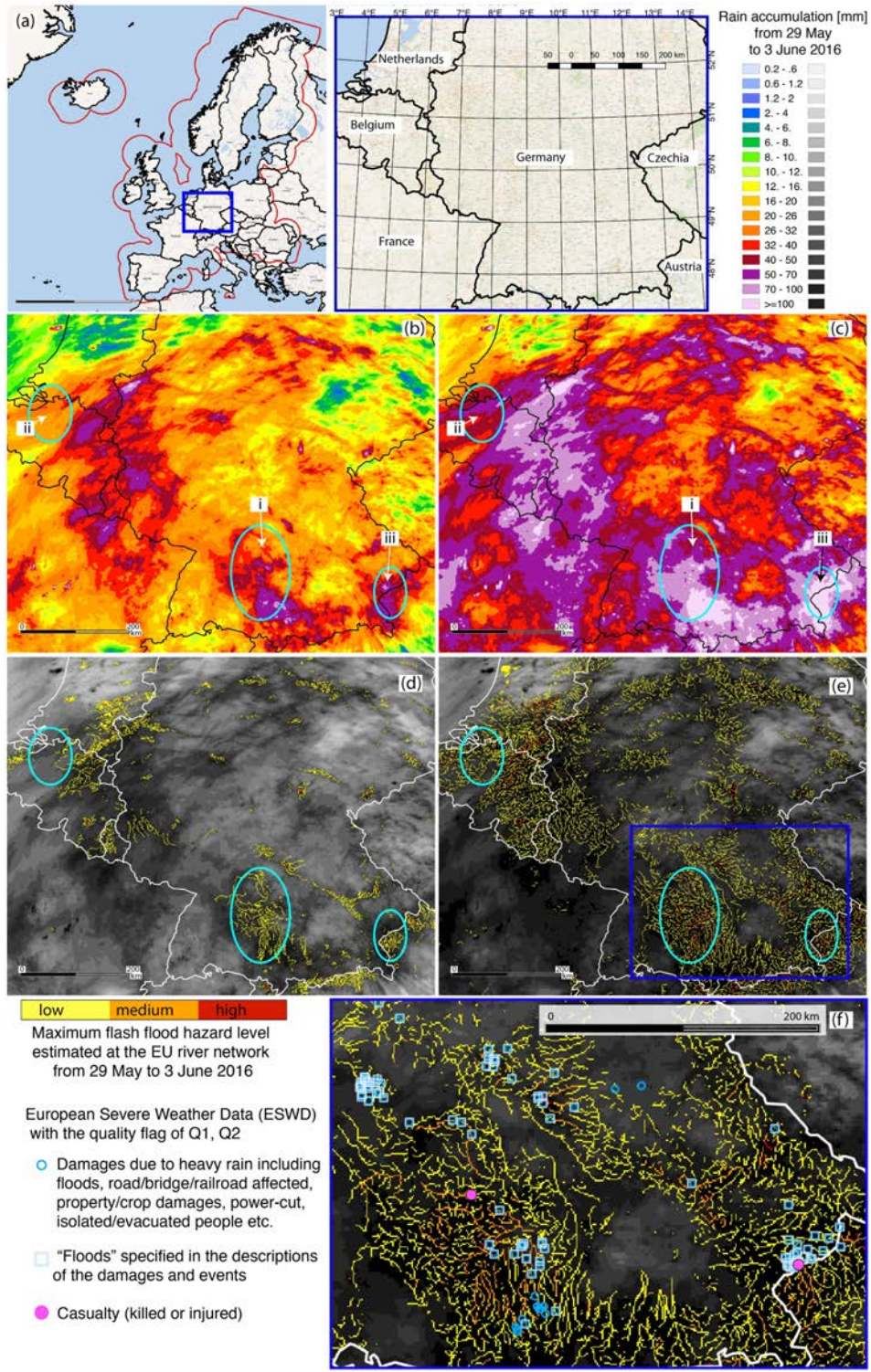

Fig. 8: An example of the impact of gauge adjustment on the flash flood hazard assessment. (a) the areas affected in Europe from 29 May to 3 June 2016. The OPERA coverage is outlined in red. (b) OPERA rain accumulations. (c) gauge-adjusted OPERA rain accumulation. A summary of the ERICHA flash flood hazard level (i.e., the maximum level extracted during the periods) obtained with (d) the OPERA accumulation inputs and (e) the gauge-adjusted OPERA rain accumulation. (f) The damage report points from ESWD plotted over the zoomed areas (blue box) of e.

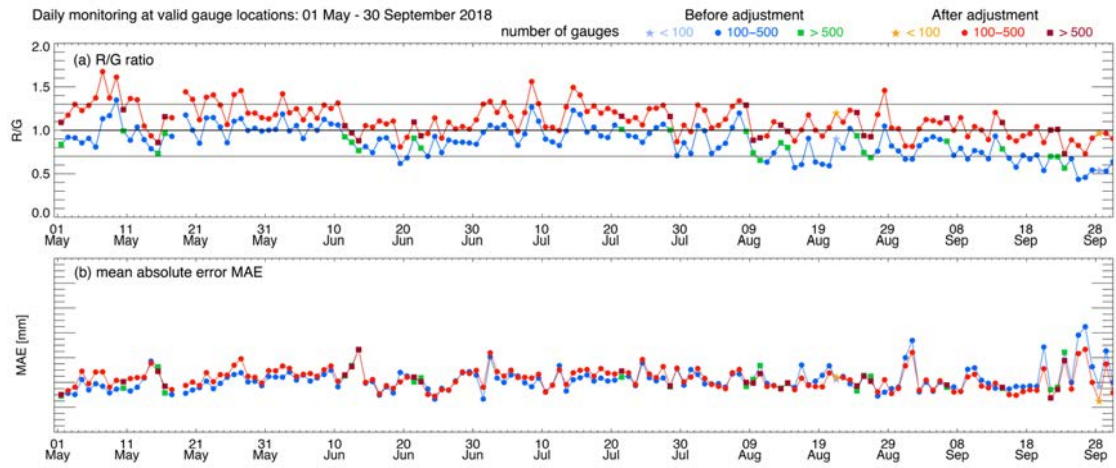

Fig. 9: Similar to Fig. 5 but (a) the ratio (R: radar-OPERA, G: gauge) and (b) mean absolute error from 1 May to 30 September 2018.